\documentclass[twocolumn, prc, amssymb, superscriptaddress, aps,
preprintnumbers,amsmath,floatfix]{revtex4}

\usepackage{multirow}
\usepackage{isotope}
\usepackage{supertabular}
\usepackage{graphicx,amsfonts,color,epsfig}
\usepackage{amsmath}
\usepackage{amssymb}
\usepackage{url}
\usepackage{xcolor}
\usepackage{tikz}
\usepackage{ulem}

\begin{document}


\title{Sensitivity of the $^{229m}$Th clock transition to the fine-structure constant in a Skyrme--Hartree--Fock--BCS approach}


\author{Nikolay \surname{Minkov}}
\email{nminkov@inrne.bas.bg} \affiliation{Institute for Nuclear Research and
Nuclear Energy, Bulgarian Academy of Sciences, Tzarigrad Road 72, BG-1784
Sofia, Bulgaria}
\affiliation{Max-Planck-Institut f\"ur Kernphysik, Saupfercheckweg 1,
D-69117 Heidelberg, Germany}

\author{Adriana P\'alffy}
\email{adriana.palffy-buss@uni-wuerzburg.de}
\affiliation{ University of W\"urzburg, Institute of Theoretical Physics and Astrophysics, Am Hubland, 97074 W\"urzburg, Germany}


\date{\today}

\begin{abstract}

The sensitivity of the $^{229m}$Th isomer transition frequency to the possible temporal variation of the fine-structure constant  $\alpha$ is investigated theoretically. We evaluate both the Coulomb as well as  the isomer energies in a selfconsistent Hartree-Fock plus Bardeen-Cooper-Schrieffer (BCS)  approach with Skyrme energy density functional,  taking a detailed account of the nuclear shape deformation and pairing correlations. Our results show that  by a fine tuning of the pairing strengths and considering axial octupole deformation, the model can predict a very low isomer energy below the keV limit, which is not accessible in the presence of imposed reflection symmetry. Furthermore, in the model solution with  octupole deformation, the difference between the Coulomb energy in the isomeric and  ground states  is one  order of magnitude larger than for the case with imposed reflection symmetry. These results allow for a fully theoretical prediction of the sensitivity of the isomer transition frequency to  $\alpha$ in the framework of a microscopic nuclear model, yielding values of $K\sim 10^3-10^4$, and for a deeper understanding of its underlying physical conditions.
\end{abstract}


\maketitle


{\it Introduction}.---
As the currently most precise timekeepers, atomic clocks can be used to search for the  variation of fundamental constants or physics beyond the standard model \cite{Safronova2018,Safronova2019}. An alternative very precise nuclear clock based on a unique nuclear transition in the $^{229}$Th isotope was proposed more than two decades ago by Peik and Tamm  \cite{Peik_Clock_2003}. The first nuclear excited state of $^{229}$Th  lies at only 8.4 eV \cite{Beck_78eV_2007,Seiferle_EnTh229m_2019, Yamaguchi_EnTh229m_2019,Sikorsky2020, TiedauPRL2024,ElwellPRL2024,ZhangNature2024},
and is a long-lived isomer with a radiative lifetime of a few $10^3$ seconds \cite{TiedauPRL2024,ElwellPRL2024,ZhangNature2024}. The unusually low transition energy renders it accessible by vacuum ultraviolet (VUV) laser technology currently under development. Broadband laser excitation of the nuclear isomer doped in VUV transparent crystals was recently achieved \cite{TiedauPRL2024,ElwellPRL2024}, followed by VUV frequency comb excitation \cite{ZhangNature2024}. Finally, the first two  solid-state nuclear clocks in Th-doped CaF$_2$ have just been reported (June 2026) \cite{thorsten2026,chinese2026}.  Combined with further progress in VUV laser technology, the nuclear clock transition promises increased sensitivity to the possible temporal variation of the fine structure constant $\alpha$ and the dimensionless ratio  between the average quark mass and the quantum chromodynamic scale  $m_q/\Lambda_{QCD}$  \cite{Flambaum06,PLB07_Hayes,JPG07_He,NPA08_He,JPG08_He,PRC08_Hayes, Berengut2009,PRC09_Litvinova_HFB,EPL09_Flambaum, APBer19_Thirolf,Flambaum2020,NRP21_Beeks,PRC22_Fadeev,Peik2021}.

Assuming that only protons contribute to the Coulomb part of the nuclear Hamiltonian and neglecting any isospin-violating contributions of the strong force, only the Coulomb energy of the nucleus depends linearly on $\alpha$ \cite{PLB07_Hayes}. Then, the relative temporal change $\delta\alpha/\alpha$ can be determined by measuring the corresponding drift $\delta\nu$ in the isomer transition frequency $\nu$ and dividing it by the change $\Delta V_{C}$ in the Coulomb energy from the isomeric state (IS) to the  ground state (GS),
$\delta\alpha/\alpha = h\delta\nu/(\Delta V_{C})$ \cite{Flambaum06,PLB07_Hayes,PRC09_Litvinova_HFB}.
The dimensionless enhancement factor $K$ defined as $  \frac{\delta\nu}{\nu}=K\frac{\delta \alpha}{\alpha}$ quantifies the sensitivity of the nuclear clock transition to $\alpha$-variation \cite{Safronova2019}. This sensitivity is related to the Coulomb energy by
$K=\Delta V_{C}/(h\nu)=\Delta V_{C}/E_{\mbox{\scriptsize IS}}$, where
 $E_{\mbox{\scriptsize IS}}=h\nu$ denotes the isomer energy and  $h$ is the Planck constant \cite{PLB07_Hayes,Berengut2009}. With current estimates of the $^{229}$Th Coulomb energy in the range of tens to hundreds of keV \cite{NPA08_He,Berengut2009,PRC09_Litvinova_HFB,Flambaum2020}, and mHz expected absolute frequency uncertainty of the nuclear clock \cite{Campbell_Clock_2012}, an improvement by several orders of magnitude of the current   atomic-clock determined   limit  $\delta\alpha/\alpha = 10^{-18}/$year \cite{PRL21_Lange}  is expected.

The main drawback  is that the most important quantity for determining $K$ and $\delta \alpha/\alpha$, the Coulomb energy difference $\Delta V_c$, is not a measurable quantity.
It can be only determined theoretically either by direct microscopic model calculations \cite{NPA08_He,PRC09_Litvinova_HFB} or through calculations of nuclear root-mean-square radii and quadrupole (and also octupole) moments of the nuclear IS and GS  \cite{Berengut2009,Flambaum2020}, with the obtained values strongly varying depending on the model.
An alternative approach combining a semi-classical prolate-spheroid model and experimentally measured quadrupole moments predicts the  value $K=5900(2300)$  \cite{Beeks2025}.
  Another issue is that the combination of $\Delta V_{C}$ with the experimental energy $E_{\mbox{\scriptsize IS}}^{\mbox{\scriptsize Exp}}$  for the calculation of $K$ does not bring much insight into the deeper nuclear structure and physical conditions which determine the sensitivity of the isomer transition to $\alpha$ variation. Attempts have been made within the relativistic mean field  approach \cite{JPG07_He,NPA08_He} and the Hartree-Fock-Bogoliubov (HFB) approximation \cite{PRC09_Litvinova_HFB} to determine $\Delta V_{C}$ for $^{229m}$Th and subsequently the sensitivity $K$. These calculations, however, raise questions due to the large span in the obtained $\Delta V_{C}$ values and the rather rough reproduction of the isomer energy, which varies from hundreds of keV \cite{PRC09_Litvinova_HFB} to values of MeV \cite{JPG07_He,NPA08_He}. A number of macroscopic \cite{Minkov_Palffy_PRL_2017,Minkov_Palffy_PRC_2021} and microscopic models \cite{hfbcs_th229,Chen2025,Zhou2025} address the isomer energy without calculating the Coulomb energy, see also a review in Ref.~\cite{Lu2026}.
  The question remains, how reliable is the Coulomb energy calculated within  models which are so far off predicting the IS energy?

In this Letter we report on a solution of the above issues through a consistent and very fine model evaluation of both isomer $E_{\mbox{\scriptsize IS}}$ and Coulomb $\Delta V_{C}$ energies in $^{229}$Th based on the interplay between nuclear shell structure, pairing correlations and deformation. To this end we employ the self-consistent Hartree-Fock plus Bardeen-Cooper-Schrieffer (HFBCS)  approach with the well-established Skyrme III energy density functional (EDF) based on its recent application to the $^{229m}$Th problem \cite{hfbcs_th229}.
Since the  extremely fine energy scale of $^{229m}$Th is very sensitive to the odd neutron which polarizes  the meanfield and suppresses the pairing correlations, in this work we do not fix the pairing parameters with respect to the measured energy of the first $2^+$ state in the even-even $^{228}$Th nucleus as in Ref.~\cite{hfbcs_th229}, but rather allow them to vary.
 We consider octupole deformation as the charge distributions in the different octupole shapes in the GS and IS are expected to induce considerable Coulomb energy difference.

We find that a moderate change in the pairing strengths reveals a region in which the  GS and  IS levels gradually approach each other and cross at a certain point. By finer bracketing the crossing area, we can localize a theoretical energy value $E_{\mbox{\scriptsize IS}}^{\mbox{\scriptsize Th}}$ in the sub-keV range closely to the experimental value of 8.4 eV, thus marking a breakthrough for the microscopic model description of the $^{229m}$Th energy. We obtain values of hundreds of keV for the Coulomb energy  $\Delta V_{C}$, smoothly and just moderately varying in the pairing strength parameter space, leading the
 enhancement factor  $K\sim 10^{3}-10^4$, providing a  fully theoretical benchmark for the sensitivity of $^{229m}$Th to the possible temporal variation of $\alpha$.


{\it Model details and application steps}.---We apply the HFBCS approximation with seniority pairing interaction and selfconsistent blocking of the odd neutron as formulated in Ref.~\cite{PRC15_hfbcs} and used in recent works \cite{PRC22_hfbcs_isomers,PRC24_hfbcs_isomers,hfbcs_th229}. The SIII Skyrme   parametrization \cite{Beiner75_SIII} is implemented with a ``minimal'' scheme of time-odd terms including spin and current vector fields (see Appendix A in Ref.~\cite{hfbcs_th229}). The algorithm is applicable with and without imposed reflection symmetry allowing for model solutions with axial (pear-shape) octupole deformation in the reflection-unconstrained version. The HF Hamiltonian is diagonalized in the basis of the axially-symmetric deformed harmonic oscillator \cite{Vautherin73}. The basis functions are specified by the parameters $b$ (spherical equivalent oscillator constant) and $q$ (deformation parameter) optimized to provide minimal total energy for the solution (see \cite{hfbcs_th229} for details). The basis is truncated at the $N_0 + 1 = 17$ major oscillator shell. The Hamiltonian matrix elements are calculated through Gauss-Hermite and Gauss-Laguerre quadratures on a grid over the axial $z$- and radial $r$- coordinates, with $NG_{z}=30$ and $NG_{r}=15$ points, respectively. These calculation conditions were adopted in the study of actinide and transactinide nuclei \cite{PRC15_hfbcs,PRC22_hfbcs_isomers,hfbcs_th229} and set a benchmark for the model features in these mass regions.
Here we use the basis parameter values $b=0.480$ and $q=1.20$ optimized in Ref.~\cite{hfbcs_th229} for the $^{229}$Th problem.

As starting point for the
constants of neutron and proton BCS pairing we use the estimates $G_n=16.5$~MeV and $G_p=14.9$~MeV determined for the even-even core $^{228}$Th in Ref.~\cite{hfbcs_th229}. In that work, it was shown  that by adjusting the BCS pairing constants to the first $2^{+}_{1}$ energy of the core nucleus $^{228}$Th, the HFBCS calculations give $E_{\mbox{\scriptsize IS}}^{\mbox{\scriptsize Th}}$ within the range of tens of keV. The two values fulfil the ratio $G_p/G_n \approx 0.9$ which was shown to be relevant for the HFBCS application in the regions from rare-earth to trans-actinide and superheavy nuclei under the currently adopted calculation conditions \cite{PRC22_hfbcs_isomers,PRC24_hfbcs_isomers}.
We take these values as a starting point in our study while keeping their ratio as a benchmark to constrain the explored areas of pairing strengths.

The isomer energy is determined as
\begin{eqnarray}
E_{\mbox{\scriptsize IS}}^{\mbox{\scriptsize Th}}=E^{*}(3/2^{+})= E_{\mbox{\scriptsize tot}} \left(3/2^{+}\right)-E_{\mbox{\scriptsize tot}}\left(5/2^{+}\right) \, ,
\label{Eiso}
\end{eqnarray}
where $E_{\mbox{\scriptsize tot}}(3/2^{+})$ and $E_{\mbox{\scriptsize tot}}(5/2^{+})$ are the calculated total energies for the blocked $K^{\pi}=3/2^{+}$ and $K^{\pi}=5/2^{+}$ orbitals, respectively. Here, $K$ and $\pi$ refer to the projection of the total nuclear angular momentum $\hat{I}$ on the body-fixed principal symmetry axis and the parity of the nuclear level, respectively.
The total energy contains the Coulomb energy contribution $E_{\mbox{\scriptsize Coul}}(K^{\pi})$. The latter is calculated through the Coulomb EDF contribution in the HFBCS Hamiltonian as explained in Ref.~\cite{PRC15_hfbcs} [see Eqs.~(5) and (A5) therein]. The Coulomb energy difference is then given by:
\begin{eqnarray}
\Delta V_{C}=E_{\mbox{\scriptsize Coul}}\left(3/2^{+}\right) -E_{\mbox{\scriptsize Coul}}\left(5/2^{+}\right)\, .
\label{DeltaEcoul}
\end{eqnarray}

\begin{figure*}[ht]
\centering
\includegraphics[width=8.5cm]{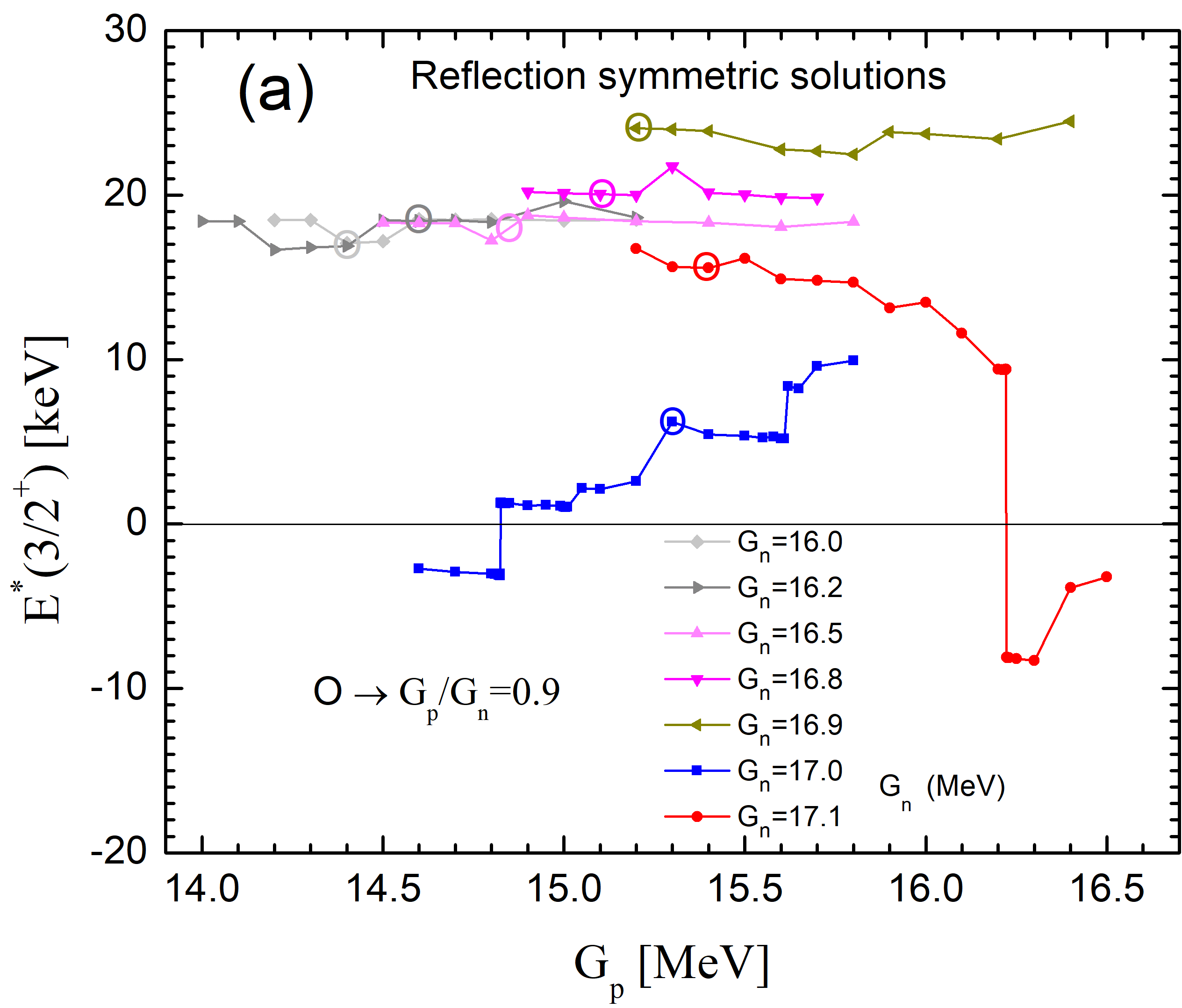}
\includegraphics[width=8.8cm]{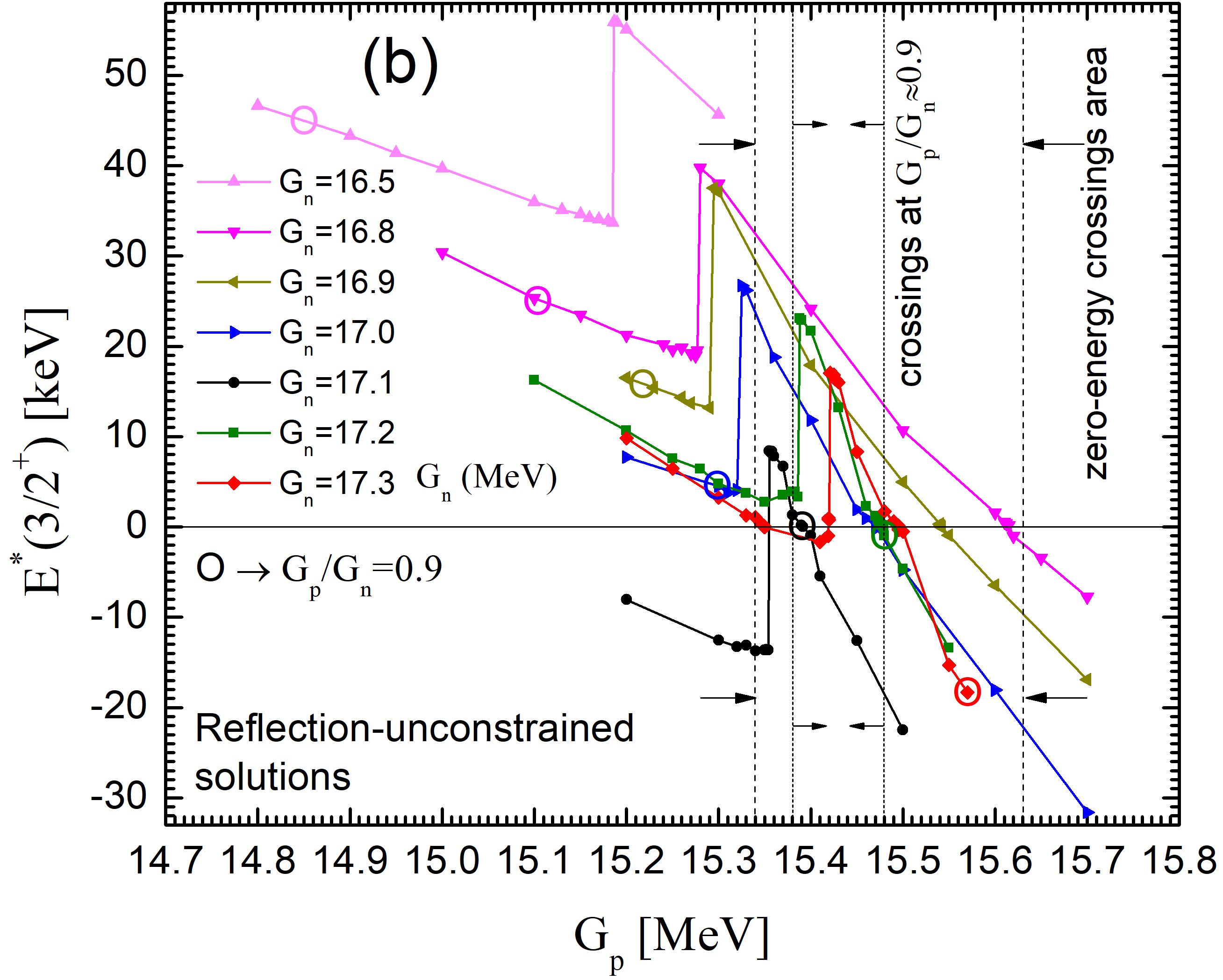}
\caption{$^{229m}$Th energy $E^{*}(3/2^{+})$, Eq.~(\ref{Eiso}), obtained in (a) reflection-symmetric and (b) reflection-unconstrained HFBCS calculations as a function of $G_p$ for several values of $G_n$. The points with $G_p/G_n=0.9$ are pinpointed by circles.}
\label{fig:eiso_gngp}
\label{fig:eiso_gngp_reflsym}
\end{figure*}


{\it Numerical results and discussion}.---We have performed HFBCS calculations with and without imposed reflection symmetry, thereby changing $G_n$ in small steps with corresponding  fine tuning of $G_p$. The pairing strengths are varied around the starting values given above in an interval that should allow us to draw definite conclusions related to the possible crossing of the GS and IS levels.  The  calculated IS energy as a function of pairing strengths for the reflection-symmetry (RS) and the reflection-unconstrained (RU) case is presented in  Figs.~\ref{fig:eiso_gngp} (a) and (b), respectively.

In the RS case, Fig.~\ref{fig:eiso_gngp}(a), the change of the neutron pairing strength in the range $G_n=16.0-16.9$ MeV yields for $E^{*}(3/2^{+})$ values between 17 keV and 25 keV with essentially flat $G_p$ dependence. This region includes the value of $E^{*}(3/2^{+})$=17.1 keV reported in Table II of Ref.~\cite{hfbcs_th229} for the set $G_n=16.0$~MeV, $G_p=14.4$~MeV. For $G_n=17.0$ MeV [blue-squares curve in Fig.~\ref{fig:eiso_gngp}(a)] the IS energy $E^{*}(3/2^{+})$  goes down with decreasing $G_p$ and reaches values $\sim$1 keV before switching with a jump at $G_p=14.83$~MeV to negative values. The latter correspond to an interchange of the $3/2^{+}$ IS with $5/2^{+}$ GS. For this energy curve the ratio $G_p/G_n=0.9$ with $G_p=15.30$~MeV corresponds to $E^{*}(3/2^{+})=6.2$ keV (blue circle). For $G_n=17.1$ MeV we get larger $E^{*}(3/2^{+})$ values in the range of $9-17$ keV with a larger jump of $\sim$18 keV to negative values at $G_p=16.20$~MeV. Overall, the RS solution does not display a smooth level crossing and is limited to keV-range IS energies.

\begin{figure}[ht]
\centering
\includegraphics[width=\linewidth]{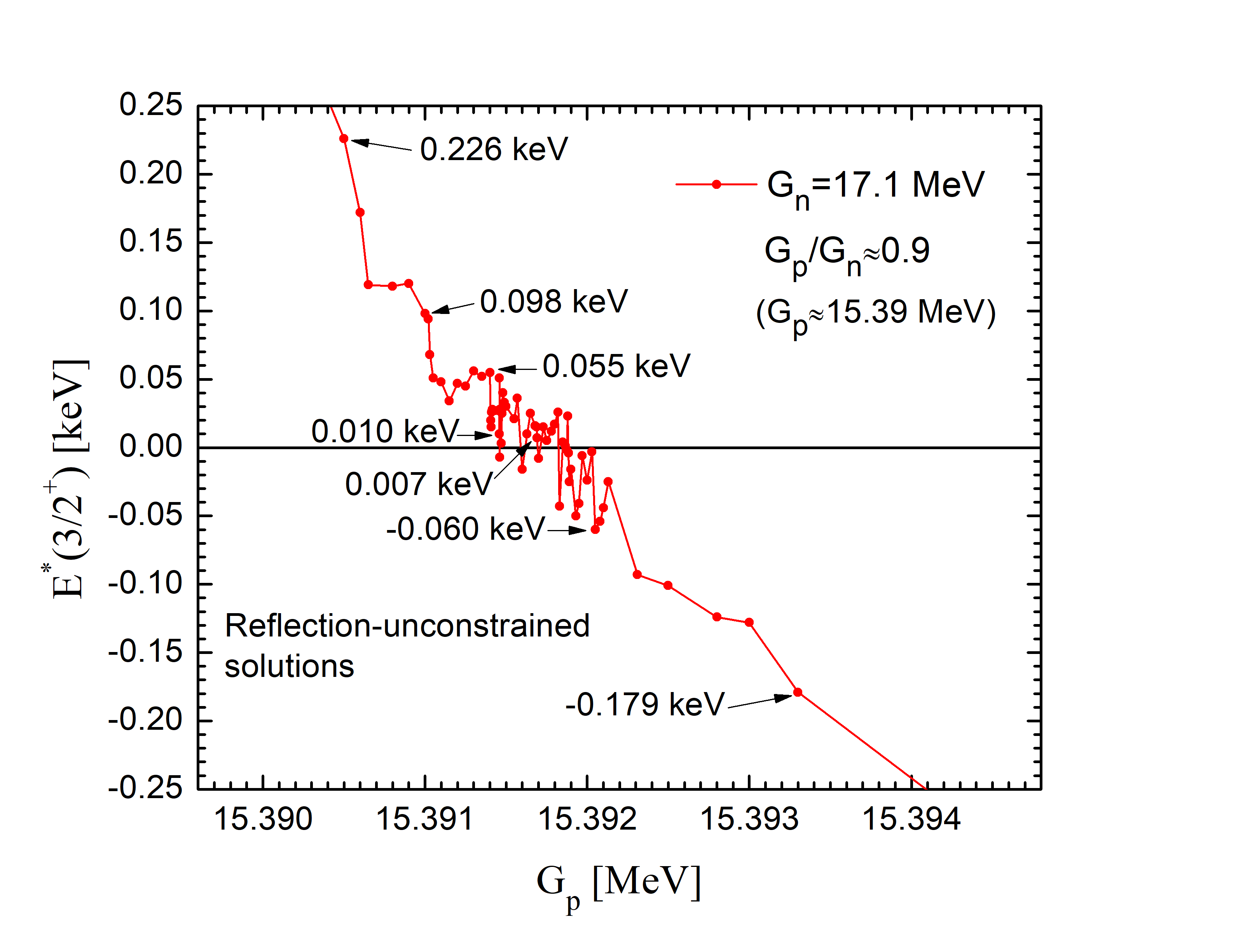}
\caption{ Zoom of $E^{*}(3/2^{+})$ energy  obtained by the RU HFBCS solution as a function of $G_p$ with $G_n=17.10$ MeV in the zero-energy crossing area.} \label{fig:eiso_zoom}
\end{figure}

\begin{figure}[ht]
\centering
\includegraphics[width=\linewidth]{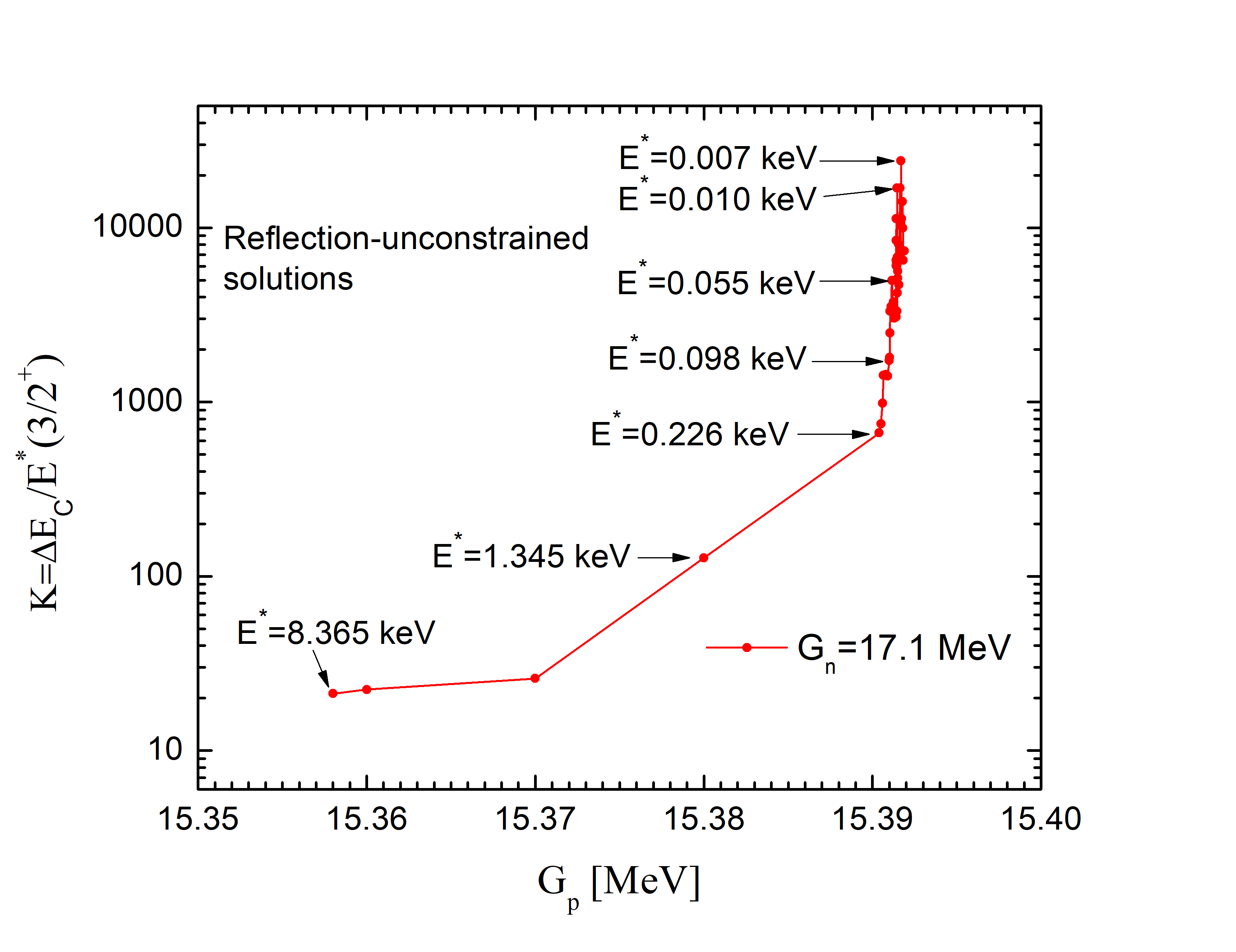}
\caption{The enhancement factor $K$ for positive $E^{*}$-values obtained by the RU HFBCS solution  as a function of $G_p$ with $G_n=17.10$ MeV in the zero-energy crossing area.} \label{fig:K}
\end{figure}

The results for the RU case are presented in Fig.~\ref{fig:eiso_gngp}(b). The change of $G_n$ in the range $G_n=16.5-17.3$ MeV provides for $E^{*}(3/2^{+})$ a family of similar mutually shifted curves depending on $G_p$. With increasing $G_p$ each curve shows a jump of about 20 keV from low to high $E^{*}(3/2^{+})$ values after which the isomer energy gradually goes down by crossing the zero energy (except for the $G_n=16.5$ MeV curve) at a certain $G_p$ value. For $G_n=17.1$ MeV [black-bullets curve in Fig.~\ref{fig:eiso_gngp}(b)] the jump in  $E^{*}(3/2^{+})$ occurs between negative and positive values, while the crossing of the zero-energy reference is found in the very close vicinity of the ratio  $G_p/G_n=0.9$ with $G_p\approx 15.39$ MeV (the black circle). Here a very fine variation of $G_p$ around the third (keV) decimal digit and beyond reveals a rather detailed picture of the zero crossing area which allows us to determine the intrinsic model accuracy limit in approaching the experimental isomer energy. This is shown in Fig.~\ref{fig:eiso_zoom}. By taking $G_p$ between 15.391 MeV and 15.392 MeV one obtains $E^{*}(3/2^{+})$ between 0.060 and $-$0.060 keV, thus narrow bracketing it within the $3/2^{+}$--$5/2^{+}$ levels-crossing area. However, this is associated with certain fluctuations of about 0.050 keV showing that in this very fine energy scale the zero-crossing is not as smooth as it looks in Fig.~\ref{fig:eiso_gngp}(b). Within this area several $E^{*}(3/2^{+})$ values close to 8 eV can be obtained. The above result suggests that the numerical accuracy of our calculation reaches an intrinsic limit of about one order above the experimental isomer energy.
Although remaining beyond the current measurements precision \cite{TiedauPRL2024,ElwellPRL2024,ZhangNature2024} this accuracy is unprecedented for a nuclear structure model. It allows us to rather precisely evaluate the $3/2^{+}$--$5/2^{+}$ levels-crossing effect as a condition for the $^{229m}$Th isomer formation and theoretically assess the sensitivity of the isomer energy to the variation of $\alpha$ as illustrated in Fig.~\ref{fig:K} (to be discussed below). Compared to the RS case considered above, we find that the presence of octupole  deformation is decisive in the microscopical description of the isomer formation.

In Fig.~\ref{fig:eiso_gngp}(b) we see that the $G_p$ curves at $G_n=$16.8, 16.9, 17.0, 17.2 and 17.3 MeV  cross (in the meaning explained above) the zero $E^{*}(3/2^{+})$ energy reference for $G_p/G_n$ ratios different from 0.9 with that for $G_n=$17.2 MeV being closest to it. All together they determine an overall range of pairing strengths for which the RU HFBCS solution reaches the sub-keV scale of $^{229m}$Th. Further  detailed accuracy checks (some of them discussed below) or tests of different realizations and/or versions of the HFBCS algorithm and EDFs  may suggest different positions for the actual levels-crossing point. However, the important result is that in all cases this point (or area) exists and essentially determines the $^{229m}$Th properties including its sensitivity to the temporal variation of $\alpha$.

A note is due here concerning the jumps in $E^{*}(3/2^{+})$ observed in Figs.~\ref{fig:eiso_gngp}(a) and (b).
We have checked that the single-particle (s.p.) spectra and wave-function structures in the HFBCS solutions, as well as the obtained values of deformations and electromagnetic moments, do not show any unusual behavior around the energy jump points. The same was verified for the $^{228}$Th core solutions with the same pairing constants. By inspecting the numerical contributions of the different EDF terms in the total energy---see Eqs.~(A5) and (5) in Ref.~\cite{PRC15_hfbcs}---we have found that the jumps are essentially generated by the kinetic and volume components of the Skyrme EDF. At the same time the surface, spin-orbit and time-odd components remain practically unaffected, while the pairing and Coulomb contributions occasionally feel the jump points. We thus conclude that the observed model dependence of the $E^{*}(3/2^{+})$ isomer energy on the pairing strengths is not a simple effect manifesting on a s.p. level, but rather an overall bulk effect of the self-consistent solution.

To further examine the accuracy of our approach and assess the reliability of our conclusions, we have performed an additional numerical check: we increased the number of integration points to $NG_{z}=40$ and $NG_{r}=20$ and repeated the calculations for $G_n=17.0$ MeV and $17.1$ MeV in the RS case and $G_n=17.0$--17.3 MeV in the RU case. We found that: (i) The calculations reproduce the same shapes of the curves in Fig.~\ref{fig:eiso_gngp} but shifted upwards by few tens of keV and right towards higher $G_p$ by about 0.5 MeV. The main effect of this shift is that in the RU case the $E^{*}(3/2^{+})$ zero-crossing points appear at $G_p/G_n$ slightly larger than 0.9 (e.g. 0.93 for $G_n=17.1$) providing a new benchmark for the model performance at the above grid condition; (ii) The numerical accuracy in the $3/2^{+}$--$5/2^{+}$ levels-crossing areas remains the same; (iii) The $E^{*}(3/2^{+})$ jumps observed in Fig.~\ref{fig:eiso_gngp}(a) and (b) appear with the same shapes but shifted according to the explanation in (i). This numerical check demonstrates the capability of the model algorithm to localize the relevant $3/2^{+}$--$5/2^{+}$ levels-crossing area under different numerical conditions, which is a prerequisite for its reliability.

\begin{figure*}
\centering
\includegraphics[width=8.5cm]{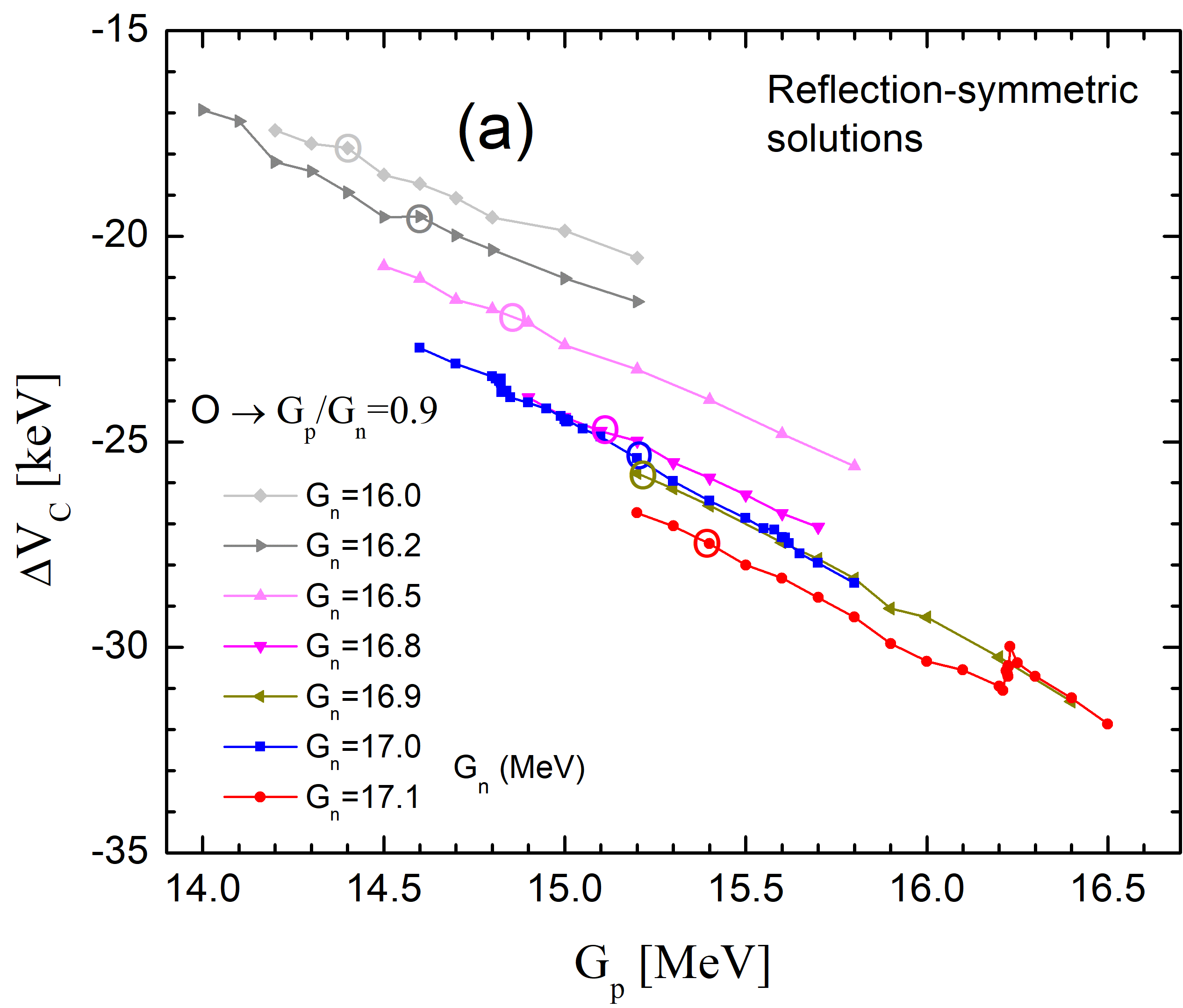}
\includegraphics[width=8.8cm]{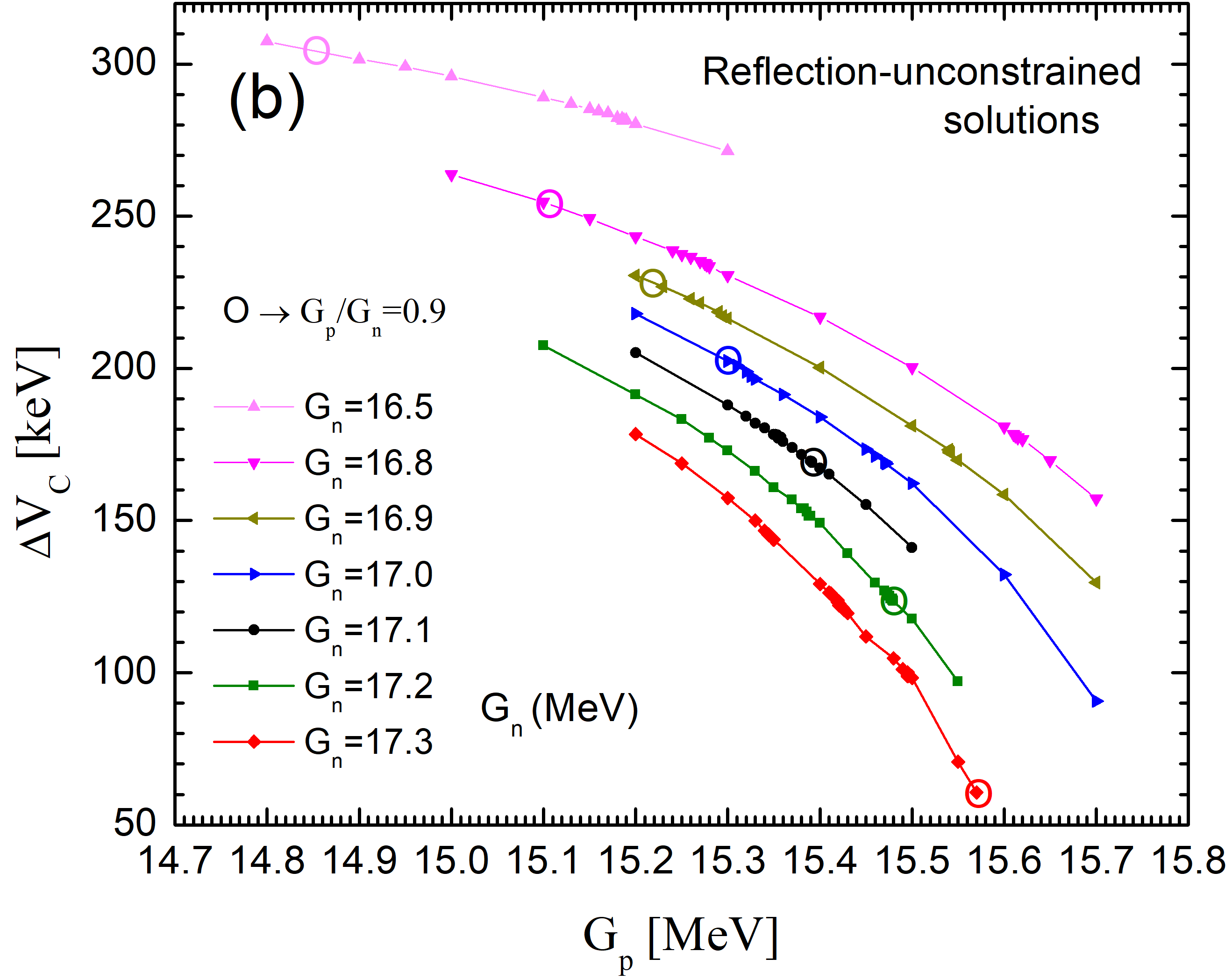}
\caption{The Coulomb energy difference $\Delta V_{C}$, Eq.~\eqref{DeltaEcoul}, between the IS and GS of $^{229}$Th obtained in (a) reflection-symmetric and (b) reflection-unconstrained HFBCS calculations as a function of $G_p$ for several values of $G_n$.
Note the different energy scales.}
\label{fig:dcoul_gngp}
\end{figure*}

Our results for the difference in Coulomb energy  $\Delta V_{C}$ in Eq.~(\ref{DeltaEcoul}) obtained in the RS and RU cases
are presented in  Figs.~\ref{fig:dcoul_gngp}(a) and (b), respectively. In the RS case, Fig.~\ref{fig:dcoul_gngp}(a), $\Delta V_{C}$ is negative and its absolute values increase linearly with the increase of $G_n$ and $G_p$ in the range between 17 keV and 32 keV. In contrast, in the RU case, Fig.~\ref{fig:dcoul_gngp}(b), $\Delta V_{C}$ is positive and decreases with $G_n$ and $G_p$ in the range from $\Delta V_{C}\approx 300$ keV to $\Delta V_{C}\approx 50$ keV.
Thus, the release of the reflection symmetry in the HFBCS solution leads to a significant change in the Coulomb energy with a considerable increase in $\Delta V_{C}$, of about one order of magnitude in absolute value. This can be explained as the result of a considerable redistribution of the charge and appearance of polarized electric dipole moment due to the axial (pear-like) octupole shape \cite{DorsoNPA86,BN96}. Also, the difference in  $\Delta V_{C}$ does not feel the jumps observed in the isomer energy $E^{*}(3/2^{+})$ as a function of $G_p$, with the exception of the point $G_n=17.1$~MeV, $G_p=16.23$~MeV in the RS case at which a relatively small jump is observed in Fig.~\ref{fig:dcoul_gngp}(a).

The sensitivity $K$ determined as the ratio of the calculated  HFBCS Coulomb and IS energies is presented in Fig.~\ref{fig:K} for the $E^{*}(3/2^{+})$ energies given in Fig.~\ref{fig:eiso_zoom}.
Since within the tiny range of $G_p=15.38-15.39$ MeV the $\Delta V_{C}$  values are $\approx 170$ keV and $E^{*}(3/2^{+})$ drops towards zero, the $K$-factor in Fig.~\ref{fig:K} sharply raises towards values of $\sim 10^{4}$ and even more. However, as shown in Fig.~\ref{fig:eiso_zoom}(a), the $E^{*}(3/2^{+})$ values obtained below 0.100 keV reach the limits of our numerical accuracy. Then, in a more conservative model prediction, the lower limit of the enhancement factor can be set to the order of $K\sim 10^{3}$.

Finally, our calculations allow to assess the shape and electromagnetic characteristics in the GS and IS of $^{229}$Th as a function of the pairing strengths and, in particular, around the $3/2^{+}$--$5/2^{+}$ levels-crossings in the RU case. We find that the electric quadrupole and octupole moments, the respective deformations and the magnetic dipole moments show a smooth behavior as functions of the pairing strength parameters  both in the RS and RU calculations without any signs of the jumps observed in Fig.~\ref{fig:eiso_gngp}.  In the levels-crossing area around $G_n=17.1$ MeV and $G_p\approx 15.39$ MeV, in the RU case, the obtained GS quadrupole deformation is $\beta_{2}=0.214$, while the octupole deformation is $\beta_{3}=0.075$. For the magnetic moments we obtain $\mu_{GS} = 0.382\mu_{N}$ and $\mu_{IS}=-0.256\mu_{N}$ which are in good agreement with the corresponding experimental values of $0.360(7)\mu_{N}$ \cite{Safronova13} and $-0.37(6)\mu_{N}$  \cite{Thielking2018,Mueller18}. A comparison of our predictions with recent estimates  made for the magnetic and electric moments of $^{229}$Th within a multi-reference DFT approach based on several Skyrme functionals \cite{ARX26_Dobaczewski_DFT} also highlights the relevance of our approach. The magnetic moment results suggest that further constraining of the overall IS-GS levels-crossing area in Fig.~\ref{fig:eiso_gngp}(b) based on the experimental  $\mu_{GS}$ and  $\mu_{IS}$ values could the subject of future work.

{\it Conclusion.}---We found that the Skyrme HFBCS solution with allowed pear-shape deformation is capable to approach unexpectedly closely the experimental $^{229m}$Th energy of $\sim$8 keV. This is due to the narrowly localized $3/2^{+}$--$5/2^{+}$ levels-crossing area in the pairing strength parameter space. Within this area the theoretical isomer energy rapidly descends into the sub-keV scale while the Coulomb energy difference remains almost constant with a value exceeding hundred keV. With the identified model accuracy, reaching the energy limit of 0.1 keV, this condition safely determines the sensitivity to the variation of $\alpha$ on the order of $10^{3}$. The finer tuning of the solution below this limit can enhance this value to $\sim 10^4$. These values are in good agreement with the recent result of Ref.~\cite{Beeks2025}.
Our  results open the way for an unprecedented precise application of the HFBCS approach to the $^{229m}$Th phenomenon and highlights the relevance of the microscopic EDF theory in the study of extremely fine nuclear structure properties and their relation to fundamental physical constants and phenomena.

The authors thank L. Bonneau for fruitful discussions. This work is supported by the Bulgarian National Science Fund (BNSF) under Contract No. KP-06-N98/2. AP gratefully acknowledges the Heisenberg Program of the
Deutsche Forschungsgemeinschaft (DFG, German Research Foundation), project-id 435041839.

\bibliography{refs}

\end{document}